# Three-dimensional angular deviation and diffraction efficiency of a grating in Littrow-configuration ECDL

Biao Chen[a,b], Y. Liu[a,c,*], Daping He[b], He Chen[d,c], Kaikai Huang[c,*] and Xuanhui Lu[c]

[a]*School of Physics, Hubei University, Wuhan, 430062*
[b]*Hubei Engineering Research Center of RF-Microwave Technology and Application,Wuhan University of Technology, Wuhan, 430070*
[c]*Institute of Optics, Department of Physics, Yuquan Campus, Zhejiang University, Hangzhou, 310027*
[d]*School of Optics and Photonics, Beijing Institute of Technology, Beijing, 100081*

**ABSTRACT**

We consider in this paper the angular deviation and diffraction efficiency of the reflection gratings in Littrow-configuration for applications of external cavity diode laser using the rigorous coupled-wave analysis method. We consider the three-dimensional diffraction case in general, where the incidence plane is un-parallel with the grating vector, i.e. conical diffraction. The angular tolerance of arbitrary gratings under plane and conical diffraction are thus derived and presented. A typical blazed grating is chosen as an example to calculate its diffraction efficiency using the rigorous coupled-wave analysis method. Furthermore, we point out that the angular tolerance and reflection efficiency can be improved if the appropriate parameter settings are selected for Littrow-configuration devices, including incidence angle, diffraction order, grating period and blazed angle. Otherwise, a tiny slanting angle of the grating vector deviated from the incidence plane will deviate the feedback light away from entering the laser-diode-chip and halt laser oscillation in the external cavity. Finally, a general rule for the parameter settings in Littrow-configuration is provided as a benchmark.

## 1. Introduction

External cavity diode laser, due to its low cost and great tuneability over a wide spectral range [1], has been extensively applied in experiments and measurements on optical and atomic physics [2, 3, 4]. Littrow [5, 6, 7] and Littman configurations[8] are two typical experimental setups to achieve this tunability. In the Littrow configuration, as shown in Fig. 1, the 0 order diffraction (geometrical reflection) from the grating constitutes the output beam and the 1st-order diffraction is coupled back into the diode laser. Thus the external feedback light dominates in the mode competition, and the external cavity lasing is built. However, if due to misalignment of positioning the grating, the grating vector may not locate in the incidence plane of light, the external cavity lasing will then cease to occur. This misaligned situation results in conical diffraction, which requires the vector diffraction theory instead of scalar diffraction theory to treat the issue. The angular deviation caused by the inevitable small misalignment in experiments should be analyzed rigorously and characterized quantitatively for better applications. Also, the essential diffraction efficiency of gratings need to be considered and discussed simultaneously.

We note that this is not the conventional case of planar diffraction where the conventional one-dimensional grating equation applies, as the grating vector does not lie exactly in the plane of incidence. For the gratings of arbitrary profiles in the interface [9], the light diffraction has been investigated by the integral-equation methods [10, 11] and the differential-equation methods [12]. For the differential-equation methods, both coupled-wave analysis and normal mode approach, or namely modal approach, are equivalent [13]. Rigorous coupled-wave analysis (RCWA) method is a straightforward and fundamentally stable technique for the accurate analysis of diffraction property of electromagnetic waves with periodic structures [14]. It has been widely utilized to analyze planar [15, 16, 17], binary [18, 19, 20], arbitrary profiled [21] gratings and other periodic structures such as: meta-surface [22, 23, 24] and photonic crystal [25, 26]. The RCWA algorithm is computationally inexpensive and easy to customize so that makes it commonly used in connection with optimization algorithms [27] and neural networks [28] to optimize structures. Here we shall use the RCWA method to treat the angluar deviation and efficiency problem in Littrow-configuration external cavity diode laser (ECDL) for three-dimensional diffraction case.

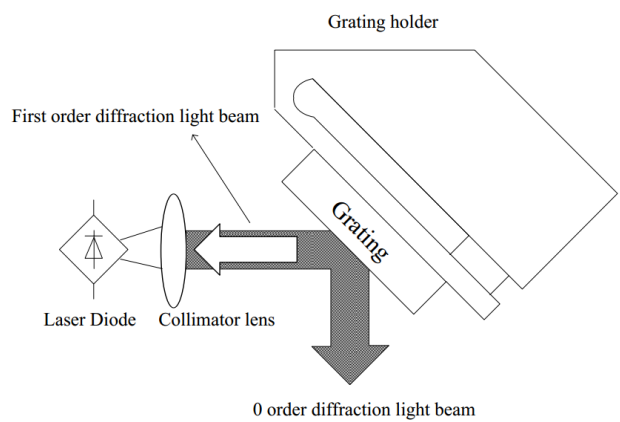

**Figure 1:** Schematic diagram of first-order Littrow-configuration External Cavity Diode Laser.

*Corresponding author: Y. Liu
✉ yangjie@hubu.edu.cn (Y. Liu); huangkaikai@zju.edu.cn (K. Huang)
ORCID(s):

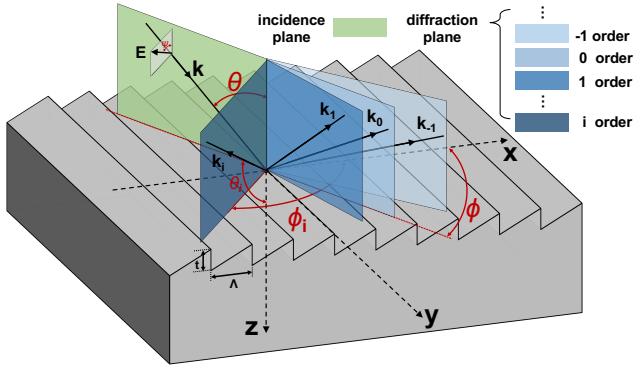

**Figure 2:** Schematic illustration for the three-dimensional diffraction problem.

In this paper, we rigorously derive the angular deviation of arbitrary gratings in three-dimensional diffraction case and give a quantitative analysis of its dependence on the related parameters, including the period of grating ($\Lambda$), incident direction (elevation angle $\theta$ and azimuth angle $\phi$) and diffraction order. In this case, the wave vector of diffraction will lie on a cone surface [17]. The diffraction efficiency is related to the shape of grating, so we choose the blazed grating, a typical grating in Littow-configuration ECDL, for efficiency analysis. Utilizing RCWA method to treat the issue above, we demonstrate that the parameter settings in Littrow-configuration will affect the angular deviation and efficiency to some extent, which are two key parameters determining the performance of the ECDL. In Sec. 4 of this paper, we provide a general rule for the parameter settings in Littrow-configuration devices for different applications.

## 2. Angular deviation analysis for arbitrary gratings

### 2.1. Parameter definition and grating equation in three-dimensional diffraction

We consider the conical diffraction in a one-dimensional grating with a period $\Lambda$ along $x$ axis. In the following content of this paper, $\theta$ denotes the incidence angle between the wave vector and $+z$ axis, $\phi$ the azimuthal angle between the incidence plane and $+x$ axis as shown in Fig. 2. The light emitted from laser diode is considered as linear-polarized. The polarization direction will be denoted by angle $\psi$ below. Note that all three angles of incident light are all scalar quantities that can be either positive or negative, which are positive $\theta > 0$, $\phi > 0$ in Fig. 2 along with several diffraction planes. The incidence plane (green) and diffraction planes (blue) are defined by the wave vector $\mathbf{k}$ or $\mathbf{k}_i$ and the boundary normal vector $\mathbf{z}$. The direction of $i$-th order diffracted wave is characterized by angle $\theta_i$ and $\phi_i$. The definition of diffraction angle is the same as that of incident light and labelled in Fig. 2.

Three-dimensional diffraction, i.e. conical diffraction occurs when the grating vector is un-parallel with the incidence plane. The grating vector $\mathbf{K}$ is defined by the period of grating,

$$\mathbf{K} = \frac{2\pi}{\Lambda}\mathbf{x}, \quad (1)$$

where $\Lambda$ is the periodicity of grating and $\mathbf{x}$ is the direction of the period.

Presumably, the grating vector $\mathbf{K}$ should lie parallel to the incidence plane in order to make the diffracted wave go back to laser diode as the white arrow shows in Fig. 1, i.e. *plane diffraction*. We will discuss the general conical-diffraction case when the grating vector becomes off the incidence plane. In the conical-diffraction case, the wave vector for the $i$-th order diffraction shall lie within the blue plane in Fig. 2. When the grating plane is slanted so that the incidence wave vector no longer lies in the grating plane, the grating vector will deviate from Eq. (1) if the coordinate system remains still. Instead we rotate the incidence wave vector to keep the grating vector unchanged. This rotation of incidence plane for angle $\phi$ is equivalent to that of the grating plane by an opposite angle $-\phi$. Under our definition above, the normalized incident electric field in region 1 ($z < 0$) can be written as,

$$\mathbf{E}_{\text{inc}} = \mathbf{u}e^{-jk_0n_1(x\sin\theta\cos\phi + y\sin\theta\sin\phi + z\cos\theta)}, \quad (2)$$

where $k_0 = 2\pi/\lambda_0$ is the vacuum wave number, $\lambda_0$ is the vacuum wavelength of incident light and $\mathbf{u}$ is expressed by

$$\begin{aligned}\mathbf{u} =& \mathbf{x}(\cos\psi\cos\theta\cos\phi - \sin\psi\sin\phi) \\ &+ \mathbf{y}(\cos\psi\cos\theta\sin\phi + \sin\psi\cos\phi) \\ &- \mathbf{z}(\cos\psi\sin\phi),\end{aligned}$$

in which $\mathbf{x}$, $\mathbf{y}$ and $\mathbf{z}$ are unit vectors in $x$, $y$ and $z$ directions respectively. The normalized solutions in region 1 ($z < 0$) and region 2 ($z > t$) are given by [15, 18]

$$\mathbf{E}_1 = \mathbf{E}_{\text{inc}} + \sum_i \mathbf{R}_i e^{-j(k_{xi}x + k_y y - k_{1,zi}(z+t))}, \quad (3)$$

$$\mathbf{E}_2 = \sum_i \mathbf{T}_i e^{-j(k_{xi}x + k_y y + k_{2,zi}z)}, \quad (4)$$

where

$$k_{xi} = k_0(n_1\sin\theta\cos\phi - i\frac{\lambda_0}{\Lambda}), \quad (5)$$

$$k_y = k_0 n_1 \sin\theta\sin\phi, \quad (6)$$

$$k_{1,zi} = \begin{cases} +[(k_0 n_1)^2 - k_{xi}^2 - k_y^2]^{\frac{1}{2}}, \\ \quad (k_0 n_1)^2 > k_{xi}^2 + k_y^2, \\ -j[k_{xi}^2 + k_y^2 - (k_0 n_1)^2]^{\frac{1}{2}}, \\ \quad (k_0 n_1)^2 < k_{xi}^2 + k_y^2. \end{cases} \quad (7)$$

Since reflection gratings are of interest to us, only backward-diffracted field wave number $k_{1,zi}$ is presented here. $\mathbf{R}_i$ is the normalized electric-field amplitude of the $i$-th backward-diffracted (reflected) wave in region 1, and $\mathbf{T}_i$ is the normalized electric-field amplitude of the forward diffracted (transmitted) wave in region 2.

With the wave number from Eqs. (5)-(7), we could obtain the direction of $i_{th}$ order of diffracted (reflected) wave as,

$$\theta_i = \pi - \arcsin(\frac{\sqrt{k_{xi}^2 + k_y^2}}{k_0 n_1}), \tag{8}$$

$$\phi_i = \arctan(\frac{k_y}{k_{xi}}), \tag{9}$$

where $\theta_i \in [\pi/2, \pi]$ and $\phi_i \in [0, \pi]$ for reflected waves. Please note the definition and sign of $k_{1,zi}$ in Eqs. (3) and (7). And $\theta_i$ is defined as the angle between vector $\mathbf{k}_i$ and $+\mathbf{z}$, so we have the correct angle in Eq. (8). We put the derivation of the grating equation for conical diffraction and its remark to Appendix.

## 2.2. Restriction condition for propagating mode

In the previous subsection, we obtained the direction of $i_{th}$ order diffracted wave with grating equation in three-dimensional case. Note that the derived diffraction direction ($\theta_i$ and $\phi_i$) is independent with the grating profile or type, which applies generally to all gratings with period of $\Lambda$ from three-dimensional grating equation (also cf. Appendix).

Before considering the angular deviation in conical diffraction for one grating, we should discuss how many non-evanescent diffracted orders exist. From Eq. (7), only when $(k_0 n_1)^2 > k_{xi}^2 + k_y^2$ the corresponding diffraction mode can propagate instead of dissipating in region 1, i.e. a real $k_{1,zi}$. The grating equation indicates that whether one diffraction wave of specified order would propagate or dissipate is determined by both the incident angles ($\theta, \phi$) and the relative wavelength ($\lambda_0/\Lambda$). As shown in Fig. 3, we calculate the dependence of the order of propagating mode on $\lambda_0/\Lambda$ under two different incidence angles, plane diffraction and conical diffraction. The results of two cases in Fig. 3 show a small discrepancy even with an azimuthal angle of 20°, which is large for an experimental misalignment angle. The number of propagating orders increases rapidly with $\lambda_0/\Lambda$ decreasing. In addition, the propagating modes that have the potential to travel back to the direction of incidence are of interest, i.e. a negative wave number of $k_{xi}$. We plot the maximum order of propagating mode with variation of incident angle $\theta$ and $\phi$ for three different $\lambda_0/\Lambda$ in Fig. 4. It indicates that if azimuthal angle $\phi$ is close to zero, the variation of incidence angle $\theta$ do not induce an extra high order diffraction mode with a fixed $\lambda_0/\Lambda$. And in the Littrow-configuration ECDL of interest, the misalignment of $\phi$ is usually not large in experimental. Consequently, the impact of incident angle $\theta$ and $\phi$ on the maximum order of propagating mode can be ignored in following analysis.

## 2.3. Angular deviation in Littrow-configuration

Littrow-configuration in literature generally refers to the 1st-order diffraction wave which travels back to the incidence direction. Here, we discuss a general case that all possible diffraction orders traveling back to incidence are considered. The diffraction angle of $\theta_i$ and $\phi_i$ can be calculated with Eqs. (8) and (9), respectively. Let us consider plane diffraction first, i.e. $\phi = 0$. Thus, restriction condition for Littrow-configuration is,

$$\theta_i + \theta = \pi. \tag{10}$$

This condition can be transformed into

$$k_{xi} = -k_x = -k_0 n_1 \sin\theta, \tag{11}$$

where $k_x$ is the component of the incident wave number along the grating period direction ($\mathbf{x}$ in our setup).

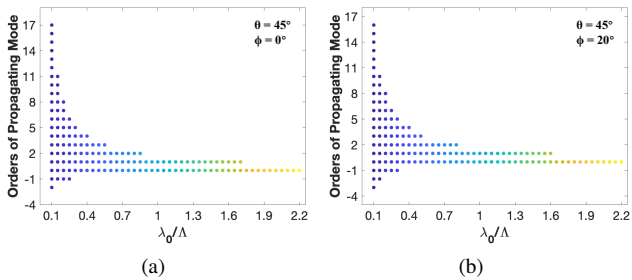

**Figure 3:** Dependence of the order of propagating mode on $\lambda_0/\Lambda$ under different incidence case. (a) $\theta$=45°, $\phi$=0° (plane diffraction); (b) $\theta$=45°, $\phi$=20° (conical diffraction).

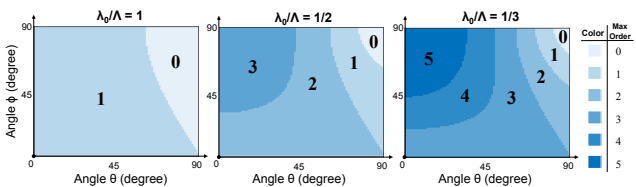

**Figure 4:** Maximum order of propagating mode with variation of incident angle $\theta$ and $\phi$ for different three $\lambda_0/\Lambda$.

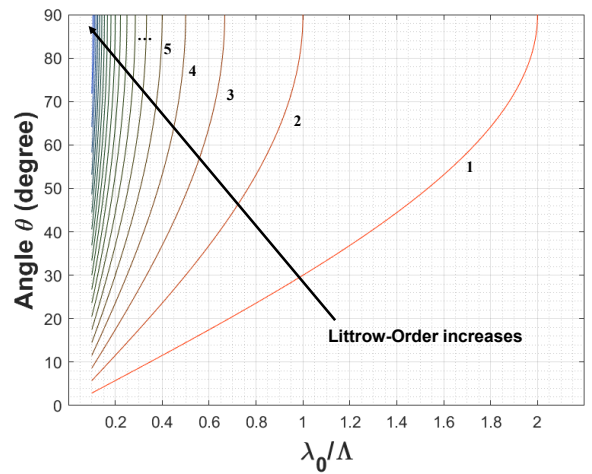

**Figure 5:** Littrow-order with variation of incidence elevation angle $\theta$ and $\lambda_0/\Lambda$.

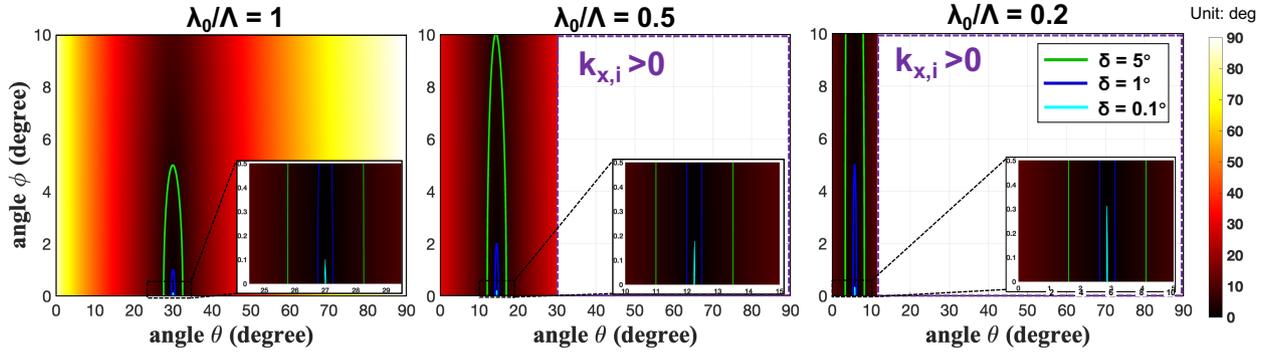

**Figure 6:** Error angle ($\delta$) of grating having different $\lambda_0/\Lambda$ with variation of incidence angle $\theta$ and $\phi$ for 1-st order diffraction, where $\theta \in [0°, 90°]$ and $\phi \in [0°, 10°]$ in this plot.

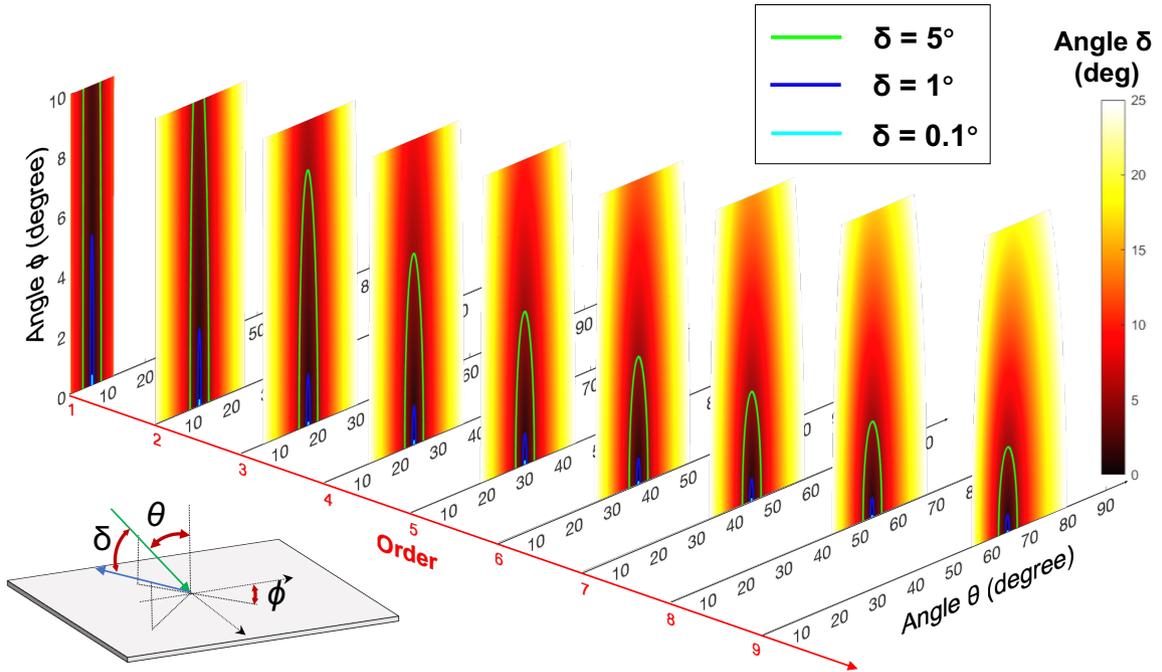

**Figure 7:** Error angle ($\delta$) of grating ($\lambda_0/\Lambda=0.2$) with variation of incidence angle $\theta$ and $\phi$ for every Littrow-order, where $\theta \in [0°, 90°]$ and $\phi \in [0°, 10°]$ in this plot.

Combining Eqs. (5) and (11) and assigning $n_1=1$, Littrow incidence condition is derived as

$$\sin\theta = i\frac{\lambda_0}{2\Lambda}, \quad (12)$$

where $i$ denotes the order of diffraction and is an integer. We name the order which satisfies Littrow condition as Littrow-order.

The Littrow condition for plane diffraction in Eq. (12) indicates that, for a fixed grating ($\lambda_0/\Lambda$), the incident angle and the diffraction order in Littrow-configuration is a one-to-one correspondence. Such a correspondence relationship is plotted in Fig. 5. From Fig. 5, we find more Littrow-orders for a grating with smaller $\lambda_0/\Lambda$. This means that a subwavelength grating, i.e. a small $\lambda_0/\Lambda$, gives rise to more high-order harmonics and thus provides more solutions in Littrow-configuration.

When it comes to conical diffraction case, the Littrow condition becomes,

$$\theta_i + \theta = \pi, \quad (13)$$
$$\phi_i - \phi = \pi. \quad (14)$$

However, considering the range of $\phi_i$ is $[0, \pi]$ and $\phi$ is a positive value in conical diffraction, the Littrow condition can never be met. The physical significance of this result is that, the angle ($\phi$) between grating vector and incidence plane provide a component of wave number along the homogenized direction of grating (**y** in our set-up). This component of incidence will transfer to that of diffraction due to the boundary condition along this direction. As a result, the wave vectors of diffraction and incidence will never parallel in the case of conical diffraction. The larger the $\phi$, the larger the $k_y$, the more severe the angle deviation.

Therefore, we need to characterize the angle deviation in conical diffraction for further discussion. We assign the angle between the wave vector of incidence (**k**) and that of $i_{th}$ order diffraction ($\mathbf{k}_i$) as error angle ($\delta$),

$$\delta = \pi - \arccos(\frac{\mathbf{k}}{k_0} \cdot \frac{\mathbf{k}_i}{k_0}), \quad (15)$$

where

$$\mathbf{k} = k_0(\mathbf{x}\sin\theta\cos\phi + \mathbf{y}\sin\theta\sin\phi + \mathbf{z}\cos\theta), \quad (16)$$

$$\mathbf{k}_i = k_0[\mathbf{x}(\sin\theta\cos\phi - i\frac{\lambda_0}{\Lambda}) + \mathbf{y}(\sin\theta\sin\phi) \\ - \mathbf{z}(\sqrt{k_0^2 - k_{xi}^2 - k_y^2}/k_0)]. \quad (17)$$

According to the results of plane diffraction, we conclude that the potential solutions for Littrow-configuration can be increased by applying a grating with a smaller $\lambda_0/\Lambda$. Hence, we compare the error angle ($\delta$) of three gratings with different $\lambda_0/\Lambda$ in Fig. 6. The data of error angle with $k_{xi} > 0$ are removed and labelled by purple dashed box in Fig. 6. This is because only the diffraction wave whose $k_{xi} < 0$ have potential to travel back to incidence position, which is of interest in this paper. The contours for $\delta = 5°$, $1°$ and $0.1°$ are depicted with yellow, green, and white color respectively, which show the tolerance of incidence angles for different $\lambda_0/\Lambda$. The insets at right bottom in each subfigure of Fig. 6 are the enlarged views around their minimum error angle. We can find that angular tolerance on $\phi$ can be improved by applying a grating with smaller $\lambda_0/\Lambda$ for 1-st order diffraction, and that on $\theta$ is not affected.

To account for this phenomena, we plot the error angle of a grating with its $\lambda_0/\Lambda = 0.2$ for every Littrow-order in Fig. 7. To represent the differences in error angle for various orders clear, the maximum value of colourbar in this plot is set as 25°. The inset at left bottom shows the definitions of three angles. From Fig. 7, for one grating with a fixed $\lambda_0/\Lambda$, the variation tendency of error angle shows that the angular tolerance of $\phi$ deteriorates with the increase of the diffraction order. Furthermore, the incident angle $\theta$ varies due to the Littrow condition for different order. If we fix the incident angle and compare their angular tolerance of $\phi$, for example, 1-st order with $\lambda_0/\Lambda = 1$ and 5-th order with $\lambda_0/\Lambda = 0.2$ whose incident $\theta$ is around 30° their angular tolerance of $\phi$ are found the same. As a result, the essential reason for that a grating with smaller $\lambda_0/\Lambda$ can obtain a larger angular tolerance of $\phi$ for 1-st order diffraction is that the incident angle $\theta$ is pushed smaller. Furthermore, the scale factor of $\lambda_0/\Lambda$ is inversely proportional to that of angular tolerance $\phi$ for same diffraction order. For example, the angular tolerance $\phi$ is doubled if we change the $\lambda_0/\Lambda$ from 1 to 0.5.

By far we have discussed the general 3D angular deviation in Littrow-configuration. In the practical scenario for Littrow-configured ECDL, the high diffraction efficiency (DE) is also a critical parameter pursued. Convectionally the blazed grating is usually adopted as the reflection grating and one uses the blazed grating's parameters to obtain a high DE once the incident angle and wavelength is determined. This convectional strategy will be referred to as *strategy 2* in the next section. Moreover in the next Sec., we will propose a new strategy referred to as *strategy 1* to determine grating parameter to gain a higher DE.

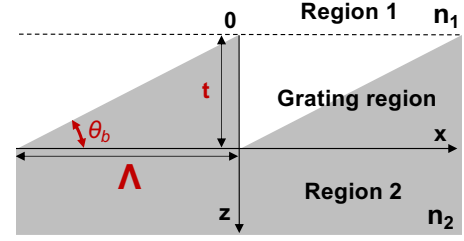

**Figure 8:** Profile of a blazed grating with refractive index $n_1$ for white region and refractive index $n_2$ for gray region.

## 3. Diffraction efficiency analysis for blazed gratings

In the previous section, we gave analysis on angular deviation due to conical diffraction for a grating and yielded change of angular tolerance with $\lambda_0/\Lambda$ and orders. For Littrow-configuration, the diffraction efficiency of gratings is another key parameter. Consequently, we apply the RCWA method to calculate the efficiency of diffraction waves in this section [18, 21, 29]. The typical grating in Littrow-configuration, blazed grating is selected here. The grating profile is shown in Fig. 8 and $n_1$ is set as 1. The wavelength $\lambda_0$ is set as 1μm for convenience. The grating material is chosen as gold and its complex refractive index $n_2$= 0.22769 - $j$·6.4731, $j = \sqrt{-1}$ at the incident wavelength of 1μm[30]. The number of harmonics is 51 in our calculations.

As shown in Fig. 9, the diffraction efficiency (DE) is calculated with RCWA method. Once $\lambda_0/\Lambda$ is determined, we can find the corresponding incident angle $\theta$ for Littrow-configuration in Fig. 5. Once $\lambda_0/\Lambda$ and thickness $t$ is determined, it means the profile of blazed grating is fixed, and thus the DE can be calculated for each pixel in Fig. 9. The blank area in the right three subfigures is because we cannot find an incident angle satisfying Littrow-configuration with such a $\lambda_0/\Lambda$ in this order. We find different grating settings ($\Lambda$ and $t$) for reaching the maximum DE under different orders in Fig. 9. Thus, we extract the maximum achieved DE and the corresponding order from Fig. 9 and demonstrate them in Fig. 10. We name this method to find the maximum DE as *strategy 1*. The maximum DE in left subfigure in Fig. 10 is 0.9740 and the corresponding settings are: $\lambda_0/\Lambda = 0.8$, thickness $t = 0.75$, order = 1, and incident angle $\theta = 23.578°$. Note that the thickness is limited in the range of 0.1 to 2.0 in this plot, so the maximum DE of high-order diffraction might not be reached under such a limit.

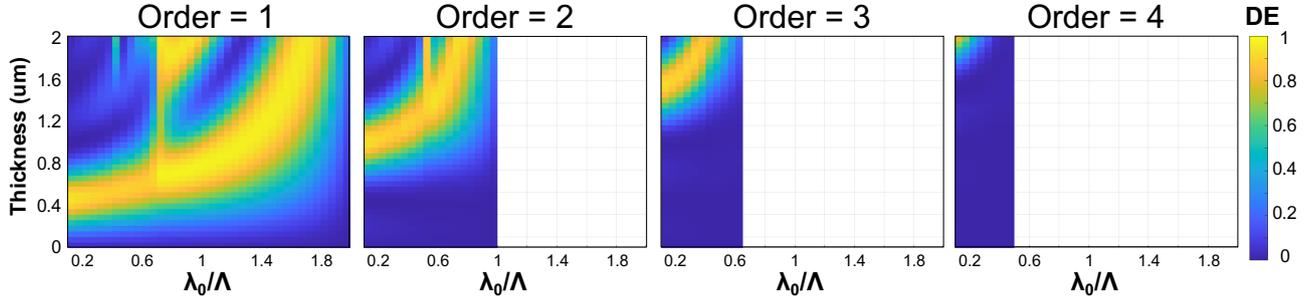

**Figure 9:** Calculated diffraction efficiency (DE) of $i_{th}$ diffraction order with the variation of $\lambda_0/\Lambda$ and thickness $t$ using RCWA method, where $\lambda_0$ is fixed at 1μm in this plot.

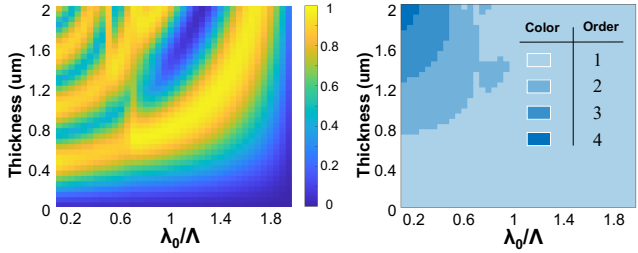

**Figure 10:** (Left) Maximum DE available extracted from Fig. 9 with the variation of $\lambda_0/\Lambda$ and thickness $t$. (Right) the corresponding order.

**Table 1**
Maximum DE for first four orders and the corresponding settings with *strategy 1* (upper columns) and *strategy 2* (lower columns).

| Order | $\lambda_0/\Lambda$ | $\Lambda$ (μm) | $t$(μm) | $\theta_b$ | $\theta$ | DE(%) |
|---|---|---|---|---|---|---|
| 1 | 0.8 | 1.25 | 0.75 | 30.96° | 23.58° | **97.40** |
|   | 0.1 | 10 | 0.5006 |  | 2.87° | 93.88 |
| 2 | 0.75 | 1.33 | 2.15 | 58.26° | 48.59° | **96.37** |
|   | 0.1 | 10 | 1.005 |  | 5.74° | 92.37 |
| 3 | 0.55 | 1.81 | 3.6 | 63.31° | 55.59° | **95.33** |
|   | 0.1 | 10 | 1.517 |  | 8.63° | 91.58 |
| 4 | 0.1 | 10 | 2.05 | 11.59° | 11.54° | **91.38** |
|   | 0.1 | 10 | 2.041 |  | 11.54° | 90.97 |

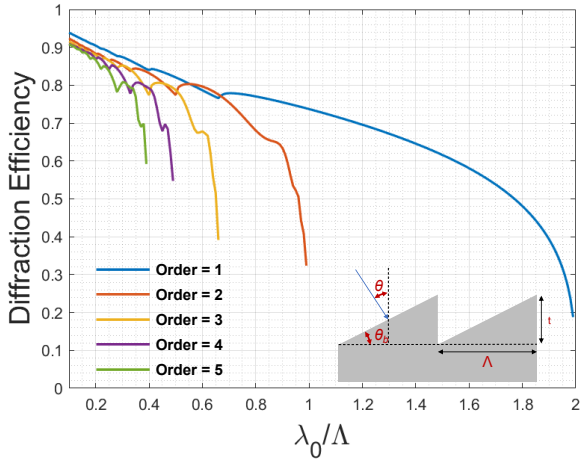

**Figure 11:** Maximum DE for different $\lambda_0/\Lambda$ and order while keeping $\theta_b$ equal to $\theta$ from *strategy 2*.

Each column in Figs. 9 and 10, i.e. a fixed $\lambda_0/\Lambda$, has the same incident angle $\theta$ obtained from Eq. (12) and the thickness varying along each column represents different blazed angle $\theta_b$ of blazed grating. It is well-known that a typical usage of blazed gratings is to keep incident angle $\theta$ equal to blazed angle $\theta_b$. Thus, we calculate the DE of a blazed grating whose $\theta_b$ is kept equal to $\theta$ by adjusting thickness $t$ in Fig. 11. We name this method to find the maximum DE as *strategy 2*. The maximum DE in Fig. 11 is 0.9388 and the corresponding settings are: $\lambda_0/\Lambda = 0.1$, thickness $t = 0.5006$, order = 1, and incident angle $\theta = \theta_b = 2.866°$.

To demonstrate the information from Figs. 9, 10 and 11 clearly, we summarize the maximum DE for first four orders in Table. 1 with two strategies shown in Fig. 10 and 11 above. Please note that the range of thickness is broadened to [0.1μm, 4μm] for *strategy 1* since this range will affect the maximum achieved DE. The results in Table. 1 indicate that a higher DE could be realized by selecting appropriate settings for one fixed incident wavelength $\lambda_0$. While the most commonly used setting that keeps $\theta$ equal to $\theta_b$ does not show any advantage on diffraction efficiency. If we need a high diffraction efficiency of gratings in applications, the *strategy 1* should be the better.

## 4. Discussion

So far we have caculated angular deviation and diffraction efficiency in Littrow-configuration. The angular deviation is analyzed for arbitrary gratings in Section 2. The results demonstrate that more high-order propagating diffraction modes can be excited by applying a grating with a smaller $\lambda_0/\Lambda$. And the incident angle $\theta$ in Littrow-configuration can be decreased and results in a larger angular tolerance in $\phi$. In experimental setup for ECDL, the angular tolerance $\tau$ can be estimated with

$$\tau = 2\arcsin\frac{D}{2L}, \quad (18)$$

where $L$ is the total external length and $D$ is the diameter of laser diode output beam. Usually angular tolerance $\tau$ is a value of $10^{-3}$ radian order (suppose $D = 100\mu m$, L = 10cm). Therefore, for this application, a grating with smaller $\lambda_0/\Lambda$ and 1-st or other low order should be chosen. Whereas for the measurement application [31], the sensitivity of Littrow-configuration is of importance. Thus a grating with smaller $\lambda_0/\Lambda$ and the highest or other high order should be chosen to obtain a small angular tolerance. However, the diffraction efficiency of gratings should be as high as possible in both two applications. For the purpose of high DE, we should apply *strategy 1* and combine the consideration on angular tolerance to find an appropriate settings, including $\Lambda$, $\theta$ and $t$.

## 5. Conclusion

In this paper, we consider and evaluate the angular deviation and diffraction efficiency of Littrow-configuration in conical diffraction case. The effects of relative wavelength and incident angles in 3D on angular deviation for any grating profile are derived and presented. Then, a blazed grating is selected and calculated in efficiency analysis by using RCWA method. Combining the results, we finally discuss a general rule for the parameter settings in Littrow-configuration under different scenarios, which should be instructive for experiments considering periodic structures in 3D scenario. Nevertheless, we believe that our grating theory can contribute to the ever-growing field of metasurface in the sense of redistribute outgoing EM waves in the spectral manner.

## Appendix: the grating equation for conical diffraction and its remark

Following Subsec. 2.1 for the wave vector of conical diffraction, we have

$$k_{xi} = k_0 n_1 \sin\theta_i \cos\phi_i, \quad (19)$$
$$k_y = k_0 n_1 \sin\theta_i \sin\phi_i, \quad (20)$$
$$k_{1,zi} = k_0 n_1 \cos\theta_i, \quad (21)$$

where $\theta_i$ is the diffraction angle of the $i$-th order diffraction wave, $\phi_i$ the azimuthal angle of the $i$-th order.

With Eqs. (19) and (20), we can eliminate $\phi_i$ and obtain,

$$k_0^2 n_1^2 \sin^2\theta_i = k_{xi}^2 + k_y^2. \quad (22)$$

Then,

$$\sin\theta_i = \frac{\sqrt{k_{xi}^2 + k_y^2}}{k_0 n_1}. \quad (23)$$

Since $\theta_i$ is defined as the angle between vector $\mathbf{k}_i$ and $+\mathbf{z}$ and the range of $\arcsin(\cdot)$ is $[-\pi/2, \pi/2]$, we have the correct angle $\theta_i$ for reflected wave as,

$$\theta_i = \pi - \arcsin(\frac{\sqrt{k_{xi}^2 + k_y^2}}{k_0 n_1}). \quad (24)$$

Similarly, with Eqs. (19) and (20), we can eliminate $\theta_i$ and obtain,

$$\phi_i = \arctan(\frac{k_y}{k_{xi}}), \quad (25)$$

where $\phi_i \in [0, \pi]$.

Furthermore, using momentum conservation along with Eqs. (19), (20), (5) and (6), we obtain the grating equation for 3-dimensional case (conical diffraction case)

$$n_1 \sin\theta_i \cos\phi_i = n_1 \sin\theta \cos\phi - i\frac{\lambda_0}{\Lambda}, \quad (26)$$
$$n_1 \sin\theta_i \sin\phi_i = n_1 \sin\theta \sin\phi. \quad (27)$$

In our problem, only the propagating wave component is of interest and otherwise evanescent wave dissipates easily in the external cavity: i.e.$(k_0 n_l)^2 > k_{xi}^2 + k_y^2$. Therefore, a limitation must be satisfied that

$$(n_1 \sin\theta\cos\phi - i\frac{\lambda_0}{\Lambda})^2 + (n_1 \sin\theta\sin\phi)^2 < n_1^2. \quad (28)$$

In such a condition, the diffraction angle is given by

$$\sin^2\theta_i = \frac{1}{n_1^2}[n_1^2\sin^2\theta - 2i\frac{\lambda_0}{\Lambda}n_1\sin\theta\cos\phi + (i\frac{\lambda_0}{\Lambda})^2]. \quad (29)$$

It is noted that the dispersion for diffraction angle $\theta_i(\phi \neq 0)$ is

$$\frac{d\theta_i}{d\lambda} = \frac{i}{n_1\Lambda\cos\theta_i} \cdot \frac{n_1\sin\theta\cos\phi - i\lambda_0/\Lambda}{[n_1^2\sin^2\theta\sin^2\phi + (n_1\sin\theta\cos\phi - i\lambda_0/\Lambda)^2]^{1/2}}. \quad (30)$$

As mentioned early in our Littrow configuration of ECDL, the first order diffraction light is expected to diffract back into the LD chip. Perfect optical feedback requires that $\theta_i = -\theta$ [32, 33]. In accordance with Eq. (1), we have Bragg condition [34]

$$2\Lambda\sin\theta = i\lambda. \quad (31)$$

## 6. Acknowledgments

L. Y. thanks Zhao Chengliang and Zhu Yingbin for their helpful suggestions. This project is supported by Ministry of Science and Technology of the People's Republic of China (2022YFC2808203), Young Scientist (NSFC11804087), Science and Technology Department of Hubei Province (2022CFB553, 2022CFA012, 2018CFB148), Department of Education in Hubei Province (T2020001, Q20211008), and Hubei University (030-017643).